\documentclass[twocolumn,aps,prb,10pt,amsmath,amssymb,superscriptaddress,nofootinbib,]{revtex4-2}

\usepackage{graphicx}
\usepackage{dcolumn}
\usepackage{bm}
\usepackage[dvipsnames]{xcolor}
\usepackage{tabu}
\usepackage{xcolor}
\usepackage{hyperref}
\usepackage{outlines}
\usepackage{multirow}
\usepackage{url}
\usepackage{physics}
\usepackage{pifont}
\usepackage{soul}

\begin{document}

\title{
Vanadium-doped HfO$_2$, multiferroic uncompromised
}
\author{Vincenzo Fiorentini}
\affiliation{Department of  Physics, University of Cagliari, Cittadella Universitaria, I-09042 Monserrato (CA), Italy}
\affiliation{Institute for Materials Science and Max Bergmann Center for Biomaterials, TU Dresden, D-01062 Dresden, 
Germany}
\author{Paola Alippi}
\affiliation{CNR-ISM, Istituto di Struttura della Materia, Area della ricerca Montelibretti, 00001 Monterotondo (RM), Italy}
\author{Gianaurelio Cuniberti}
\affiliation{Institute for Materials Science and Max Bergmann Center for Biomaterials, TU Dresden, D-01062 Dresden, 
Germany}
\date{\today} 

\begin{abstract}
Ab initio density-functional calculations show that orthorhombic $Pca2_1$ hafnia HfO$_2$ mixed with vanadium at low concentration is a ferroelectric and ferromagnetic insulator. The multiorbital degeneracy of singly-occupied V states in the nominally 4+ ionic state is broken by magnetism, reduced symmetry, and local distortion, causing a single one-electron majority state per V atom to be occupied. A gap of order 1 eV thus survives at all V concentrations, and  intrinsic polarization is preserved at the level of 60-70\% of the undoped-HfO$_2$ value. Magnetization is found to increase linearly with V content, with values of 30-40 emu/cm$^3$ at concentrations near the end of the stability range ($\sim$16\% V).
\end{abstract}

\maketitle

\section{Introduction}
Multiferroics --materials in which multiple states of  order coexist, such as ferroelectricity and some sort of magnetic order-- are rare.  Ferroelectrics that possess ferromagnetic  order  are even more rare, and usually  only weakly ferromagnetic due to e.g. Dzyaloshinskii-Moriya-type interactions. In search for a more robust sort of multiferroicity, here we explore the doping of ferroelectric hafnia with a transition metal, specifically vanadium (V). As the stoichiometry is fixed by construction, one may also view this as the mixing of two materials each endowed with one of the requisite orders: the ferromagnet is VO$_2$ in its $C 2/m$ monoclinic structure \cite{vo2,eyert} (more about this in Sec.\ref{gencons}); the ferroelectric is HfO$_2$ in the  metastable (but amenable to stabilization) orthorhombic $Pca2_1$  structure \cite{FEhafnia}. The two are mixed into Hf$_{1-x}$V$_{x}$O$_2$ (or (Hf,V)O$_2$ for short) with $x$ up to around 0.37. For this proof of concept, we  assume the starting structure of polar hafnia, and analyze the energetics and a few relevant physical quantities (signally, magnetization and polarization), using a simple model  to discuss stability against separation in bulk components. Here we do not address the upper end of the $x$ range, where a non-polar or, potentially, metallic VO$_2$ state would probably dictate the structure.
 
 The results, in a nutshell, are that polarization is conserved (though reduced) and magnetization increases linearly over the whole range considered; according to a model mixing free energy, the solid solution should be stable up to about $x$=0.15 at growth or annealing temperature 1000 K. The electronic structure evolves from the large gap of hafnia towards that of VO$_2$, with the appearance and progressive increase in weight of a V-related occupied density-of-states  feature above the upper edge of the valence band. The geometry of V configurations seems to indicate a tendency  to first form rows, then combinations of rows producing incomplete planes, then planes decorated by excess V, then V-rich regions separated by Hf-rich regions, causing the mixture to resemble a rough superlattice of hafnia and vanadia (typically stacked along the crystal directions {\bf b} and, to a lesser extent, {\bf a}). Our findings seem to compare well with a very recent set of experiments on a device stack involving the solid solution \cite{vhfo2}.

\section{Methods}

We calculate observables  from first principles within density functional theory at multiple concentrations and for multiple configurations of V substituting for Hf in ferroelectric HfO$_2$.
We use the VASP code (v.6.5) \cite{vasp} and the PAW formalism \cite{PAW} with  the datasets {\tt V}, {\tt Hf} and {\tt O$_{\tt s}$}, at a cutoff of 350 eV (1.3 times the maximum recommended cutoff).  The polarization is obtained by the Berry-phase method \cite{berry}.  

The density-functional calculations are spin-polarized GGA+U, i.e. generalized gradient approximation (with the PBE \cite{PBE} functional)  with  U corrections in the Dudarev version \cite{LDAU}. We set U--J=3 eV on V $d$ states only, and no correction on all other channels. This choice is based on preliminary comparisons with hybrid functional calculations \cite{HSE} for a few selected configurations at low V concentration, and is confirmed by previous reports \cite{V-U}. This setting  limits computational costs and makes it possible to  study a large number of configurations. Several counterchecks with hybrid functionals confirm the findings of GGA+U. The main difference is the HfO$_2$ gap, smaller (4.6 eV in GGA+U  and 5.9 in  HSE,  close to  experiment) but  still  quite sufficient to capture the evolution of the electronic structure as function of V concentration.

The calculations are done in a supercell containing 32 cations and 64 anions (a 2$\times$2$\times$2 replica of the $Pca2_1$  unit cell).  In each configuration, V atoms substitute a random sample of $n$ unique Hf atoms in the cell;  $n$ is between 1 and 12, corresponding to concentrations $x$ between 0.031 and 0.375 in steps of 0.031. A total of 1481 configurations involving 9500 vanadium atoms are considered in the averages; the pairs $(n:N_{\rm conf})$ are (2:130, 3:153, 4:165, 5:153, 6:131, 7:149, 8:145, 9:166, 10:75, 11:80, 12:80). For  $n$=1,  we use 7 configurations, finding them identical to numerical accuracy as expected since all sites are equivalent. 

The volume and internal geometry are optimized for each configuration using a 2$\times$2$\times$2 k-space grid. Observables (energy, polarization, magnetization, dielectric function, Born charges) and density of states are then computed on a 4$\times$4$\times$4 grid. (Finer grids are impractical given the relatively large  cell and  number  of configurations). A Gaussian smearing of 0.02 eV is used.
The various quantities presented below are obtained as  configurational averages with the Boltzmann weights $\exp$($E_i$/$k_BT_g$) of the  computed energies $E_i$ and the assumed growth temperature $T_g$=1000 K (well within the range generally used for hafnia \cite{growth}; see also Sec.\ref{expcomp}). The error bars in the Figures are obtained from the square root of the covariance matrix of the weighted quantities.  

\subsection{Magnetic insulator, general considerations }
\label{gencons}

For polarization to  be directly computable via the standard Berry-phase approach the system must remain insulating. If the local geometry were strictly octahedral, the excess valence electron of V would occupy a 6-fold degenerate $t_{2g}$  state. A combination of "crystal-field" and magnetic level splitting is required to remove all degeneracies and open a gap in the spectrum. For example, in  monoclinic $C 2/m$  VO$_2$  the $t_{2g}$ majority triplet splits \cite{ega1} into an $e_g^{\pi}$ doubly-degenerate state and an $a_{1g}$ singly-degenerate state, which dominates the top valence bands.  In $Pca2_1$ HfO$_2$, whose point symmetry is $mm2$, the (majority) $t_{2g}$ states are expected to split in three non-degenerate states $a_1$, $b_1$, and $b_2$, and the $e_g$ into  $a_1$ and  $b_1$. Disorder by random V substitutions, and the attendant local distortions, about which more below, indeed formally remove all symmetries. As in VO$_2$, the $a_1$ even state is expected to be lowest. Of course degeneracy removal requires a concurrent magnetic exchange splitting between the minority and majority channels, which is found to be  approximately 2 to 4 eV in the present case. The end result is thus  that a majority-channel gap (of order 1 eV) appears in the whole concentration range considered, separating occupied and empty V-derived states in the HfO$_2$ gap. 

A small percentage of the configurations are found to be ferrimagnetic, with part of the V states involved being coupled antiferromagnetically. This results in a slight reduction of the average magnetization $M$ from the ideally expected 1 $\mu_B$ per V. For example, 
the most affected case is $n$=4 where  15 \% of the configurations have $M$=2 $\mu_B$ instead of the 4 $\mu_B$ expected for a full ferromagnet, and less than 1\%  have $M$=0. ($M$ decreases by 2 $\mu_B$ for each minority state going under the Fermi level. {Note that $M$=0 can only occur if $n$ is even.}) The average magnetization per V is still a large 0.91 $\mu_B$. The effect is less frequent at other concentrations, especially at the higher end. Indeed, the magnetization averaged over all configurations is 0.957 $\mu_B$/V, close to the maximum expected value of 1 $\mu_B$.

Among our nearly 1500 configurations, we found  4 to be  metallic. Such rare cases occur at low concentration, perhaps due to V impurities being accidentally related by some approximate symmetry, and we expect them to be irrelevant in the real material.

\section{Results}

\subsection{Bulk materials, formation energy and volume}

The bulk materials are ferromagnetic VO$_2$ in the monoclinic $C 2/m$ 
 structure and ferroelectric HfO$_2$ in the  (metastable) orthorhombic $Pca2_1$ structure. The former was shown to be treatable by density functional methods by Eyert \cite{eyert}. The latter has been studied in many recent papers aiming at clarifying the nature and origin of ferroelectricity \cite{FEhafnia}{, about which we will have nothing to say. Suffice it to the present endeavor that our method does reproduce it in the polar structure. Also, we only address V behavior in the polar structure, and not in the non-polar parent(s). The polar structure does remain stable upon V incorporation in the whole range we considered.} 
 
We start from configurations found in the Materials Project \cite{mp} ({\tt mp-685097} for HfO$_2$ and {\tt mp-541404} for VO$_2$) confirming them in our setting to within the usual accuracy of 1\% or less in the lattice constants. The polarization in hafnia  points along the {\bf b} crystal axis; the computed value for the $Pca2_1$ structure supercell (referred to its  $P4_2nm$ symmetry parent) is smaller than obtained elsewhere \cite{iniguezhfo2}, possibly  due to a different reference or a relatively sparse k-sampling. This does not affect the general picture.

\begin{figure}[ht]
\centerline{\includegraphics[width=1\linewidth]{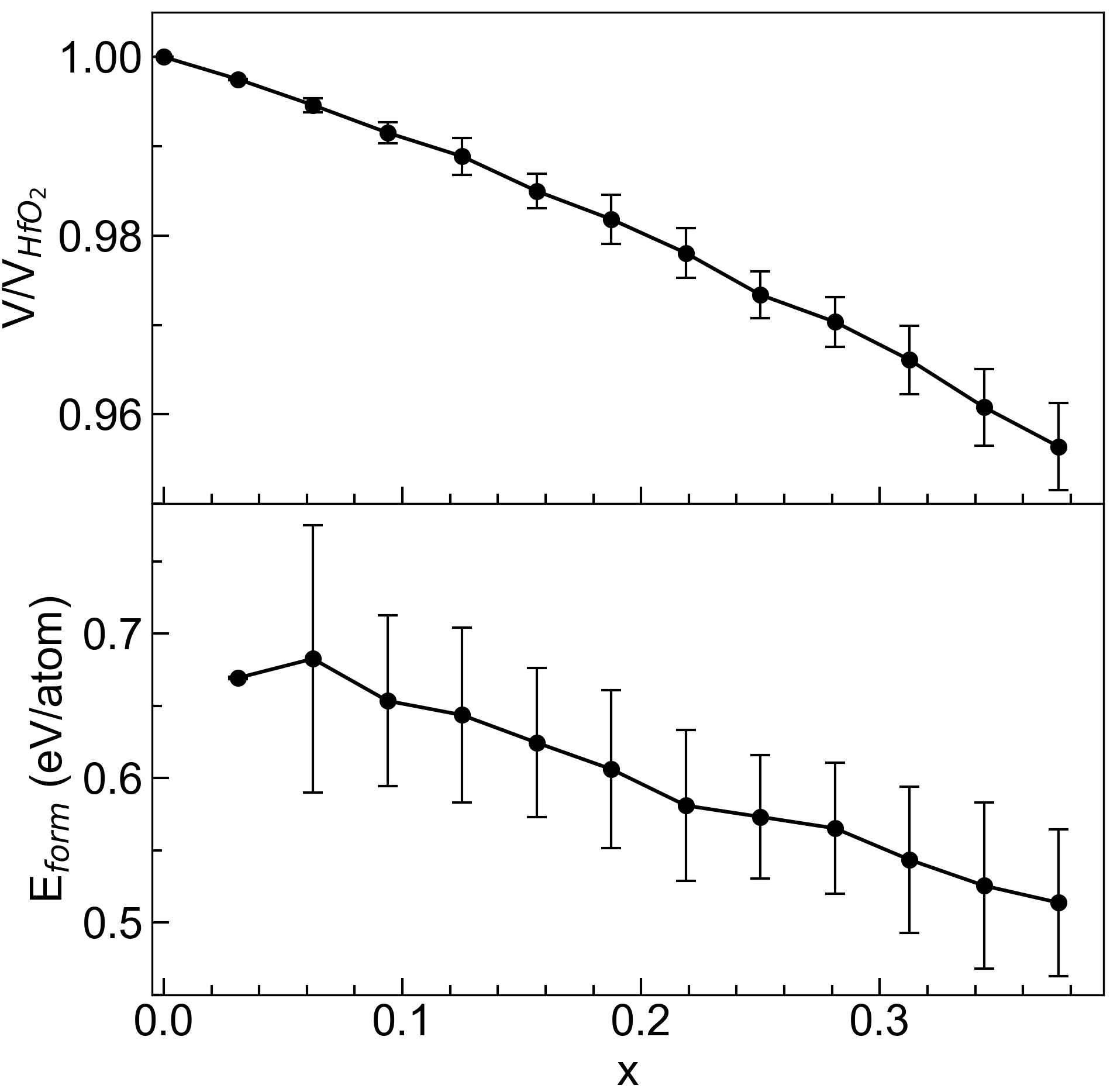}}
\caption{\label{varie1} Volume (top) and formation energy per V atom (bottom) of (Hf,V)O$_2$ vs concentration.}
\end{figure} 

We elect to study a free-standing version of the system. Epitaxial constraints, sometimes used to stabilize the ferroelectric phase, may be studied in forthcoming work. Figure \ref{varie1}, top panel, reports the volume of the simulation cell (top) vs concentration $x$. 
The volume reduction  vs $x$ is expected given the smaller volume per formula unit in VO$_2$ compared to HfO$_2$ in the same structure, although VO$_2$  in the $C 2/m$ structure has a slightly (5\%) larger specific volume. The relative variation of the three lattice constants is identical on average, i.e. the contraction is isotropic.

 The bottom panel of   Figure \ref{varie1} reports the formation energy per atom of V in HfO$_2$, obtained from the standard formula
\begin{equation}
E_f= \frac{1}{n}[(E_{\rm (Hf,V)O_2} + n \mu_{\rm HfO_2})- (E_{\rm HfO_2} + n \mu_{\rm VO_2})]  
\label{eform}
\end{equation}
with $n$ the number of V atoms, $\mu_{\rm HfO_2}$  and $\mu_{\rm VO_2}$ the energy per formula unit of bulk HfO$_2$ and VO$_2$, and  $E_{\rm (Hf,V)O_2}$, and respectively $E_{\rm HfO_2}$, the energy of the defected and undefected supercells. This yields a V concentration in HfO$_2$ in the mid 10$^{19}$ cm$^{-3}$ at 1000 K.
 
 \subsection{Mixing free energy and stability}

To determine the  stability of the solid solution we calculate its mixing free energy
\begin{eqnarray}
F_{\rm mix}&=& F(x)-(1-x) F_{\rm HfO_2} - x F_{\rm VO_2} \\
&=&E(x) - k_BT [S_m(x)+S_v(x)] \nonumber\\
& &- (1-x) [E(0) - k_BT (S_m(0)+S_v(0))]  \nonumber\\
& &-x [\mu_{\rm VO_2} - k_BT (S_m(1)+S_v(1))]\nonumber
\end{eqnarray}
where $x$=$n$/$N_{\rm Hf}$ is the fractional occupation by V on the Hf lattice, with $N_{\rm Hf}$=32 in our 96-atom simulation cell,  $E(x)$ is the same as $E_{\rm (Hf,V)O_2}$ in Eq.\ref{eform}, and $\mu_{\rm VO_2}$ is the bulk energy of $C 2/m$ VO$_2$  as in Eq.\ref{eform}. For the entropy  we adopt a standard approach \cite{mixalloy}
employed previously \cite{gaino}, whereby the mixing entropy is simply
$$ S_m(x)=-x \log x -(1-x) \log (1-x)$$ and, since the Debye temperatures $\Theta_{\rm VO_2}$=750 K and $\Theta_{\rm HfO_2}$=480 K \cite{debtemp} are both well below typical growth temperatures,  the vibrational entropy is expressed in a single-oscillator approximation as
$$
S_v(x)=3 [ (1+n_B) \log (1+n_B) - n_B \log n_B ]
$$
where $n_B$=$1/(\exp[\Theta_x/T]$--1)   
 is the Bose distribution 
and $\Theta_x=(1-x)\Theta_{\rm HfO_2}+x \Theta_{\rm VO_2}$.  
The vibrational entropy, which is not symmetric in $x$ and (1--$x$), disfavors mixing at low $x$ because the compound being admixed, VO$_2$, is stiffer and has larger Debye energy (unlike e.g. in the (In,Ga)$_2$O$_3$ case \cite{gaino}).

\begin{figure}[ht]
\centerline{\includegraphics[width=1\linewidth]{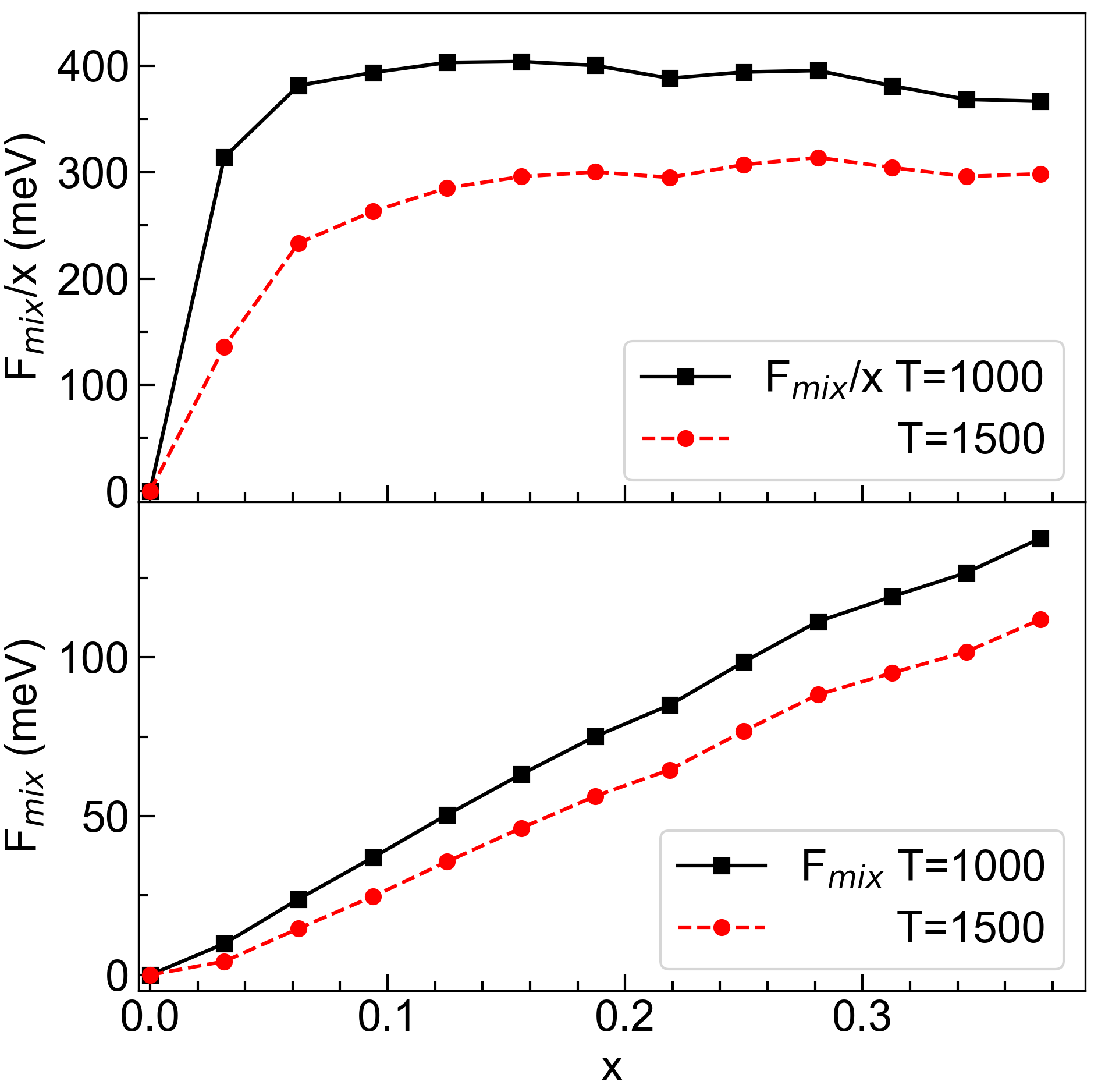}}
\caption{\label{Fmix} F$_{\rm mix}$ (bottom)  and F$_{\rm mix}$/$x$ (top) vs concentration at two temperatures. Error bars,  not shown for clarity, are of order 10\%.}
\end{figure} 

The mixing free energy is shown in Figure \ref{Fmix}, bottom panel (for the top panel see just below). The error bars, not shown to avoid clutter, are about $\pm$10\% of the plotted values. 
In order for the solid solution to be stable at a given $x$ against phase separation in the constituent bulks,  the mixing free energy should have positive local curvature \cite{mixalloy}. 
{This is hard to judge from the Figure, so we adopt  a  visual heuristic}. 
Consider $G(x)$=$F_{\rm mix}(x)/x$, shown in  Figure \ref{Fmix}, top panel. If we assume $F$$\sim$$x^{\beta(x)}$ locally in $x$, the mixture is stable in $x$ if 
$G$$\sim$$x^{\beta(x)-1}$  is upward convex {\it and} monotonically increasing. If so \cite{convex}, 1$<$$\beta(x)$$<$2 (e.g. the  case of  $G$$\sim$$\sqrt{x}$, i.e. $\beta(x)$=3/2), so that the local curvature 
$F_{xx}$$\sim$$\beta(x)(\beta(x)-1) x^{\beta(x)-2}$ is  positive. By this criterion, the stability range ends at $x$$\sim$0.16 at 1000 K, and  at $x$$\sim$0.19 at 1500 K. After that,  $G$ becomes downward convex and/or decreasing, which means  $\beta(x)$$<$1  and hence negative curvature  and instability. 
{In passing, the alternative approach of taking numerical derivatives, although plagued by noise, predicts similar stability boundaries.}

\subsection{Polarization and magnetization}

\begin{figure}[ht]
\centerline{\includegraphics[width=1\linewidth]{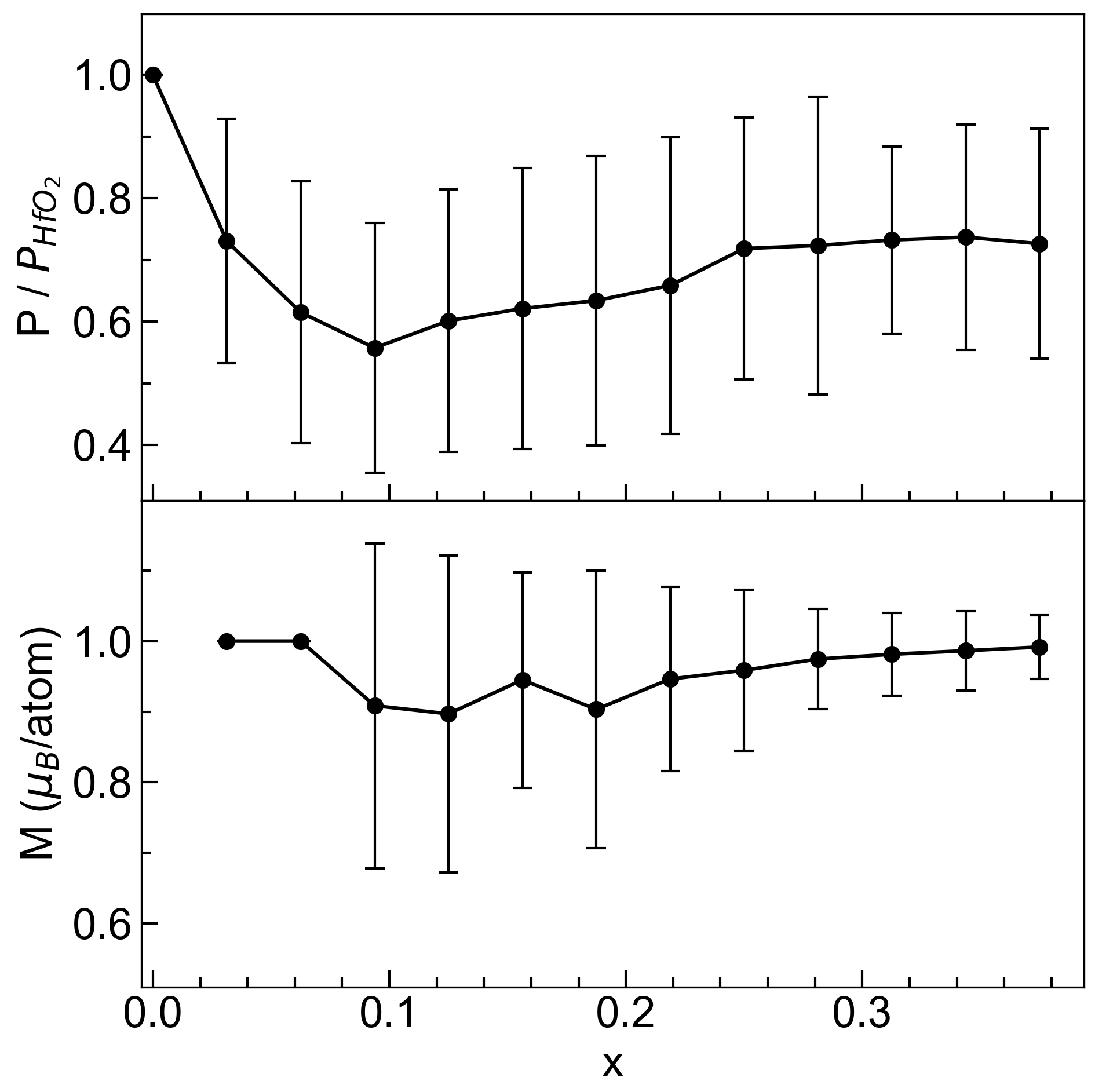}}
\caption{\label{varie2} Modulus of polarization  (top) and magnetization (bottom) vs concentration.}
\end{figure}

Figure \ref{varie2} reports the modulus $P$=$|{\bf P}|$ of the spontaneous polarization (top panel) and the magnetization $M$ (bottom panel). The magnetization is close to its ideal value of  1 $\mu_B$ per V, with downward fluctuations at low $x$ as mentioned earlier, and leveling out asymptotically to the ideal value at large $x$; the magnetization is treated as collinear, so there is no information about its direction (future work on non-collinearity and magnetoelectricity is planned).

$P$  vs. $x$ appears to saturate to about 70\% of the HfO$_2$ value. There seems to be  no systematic V-related mechanism of local dipole reduction. If V contributed less than Hf to $P$ or subtracted from it (by e.g. a weaker dynamical response or a displacement counter that of Hf in the ferroelectric pattern), $P$ should decrease with $x$.  If only mostly-HfO$_2$ regions contributed to $P$ and mostly-VO$_2$ regions did not (e.g. in the geometry of $n$=12 in Sec.\ref{structure}), $P$ should still decrease with $x$, being an extensive quantity.

\begin{figure}[h]
\centerline{\includegraphics[width=1\linewidth]{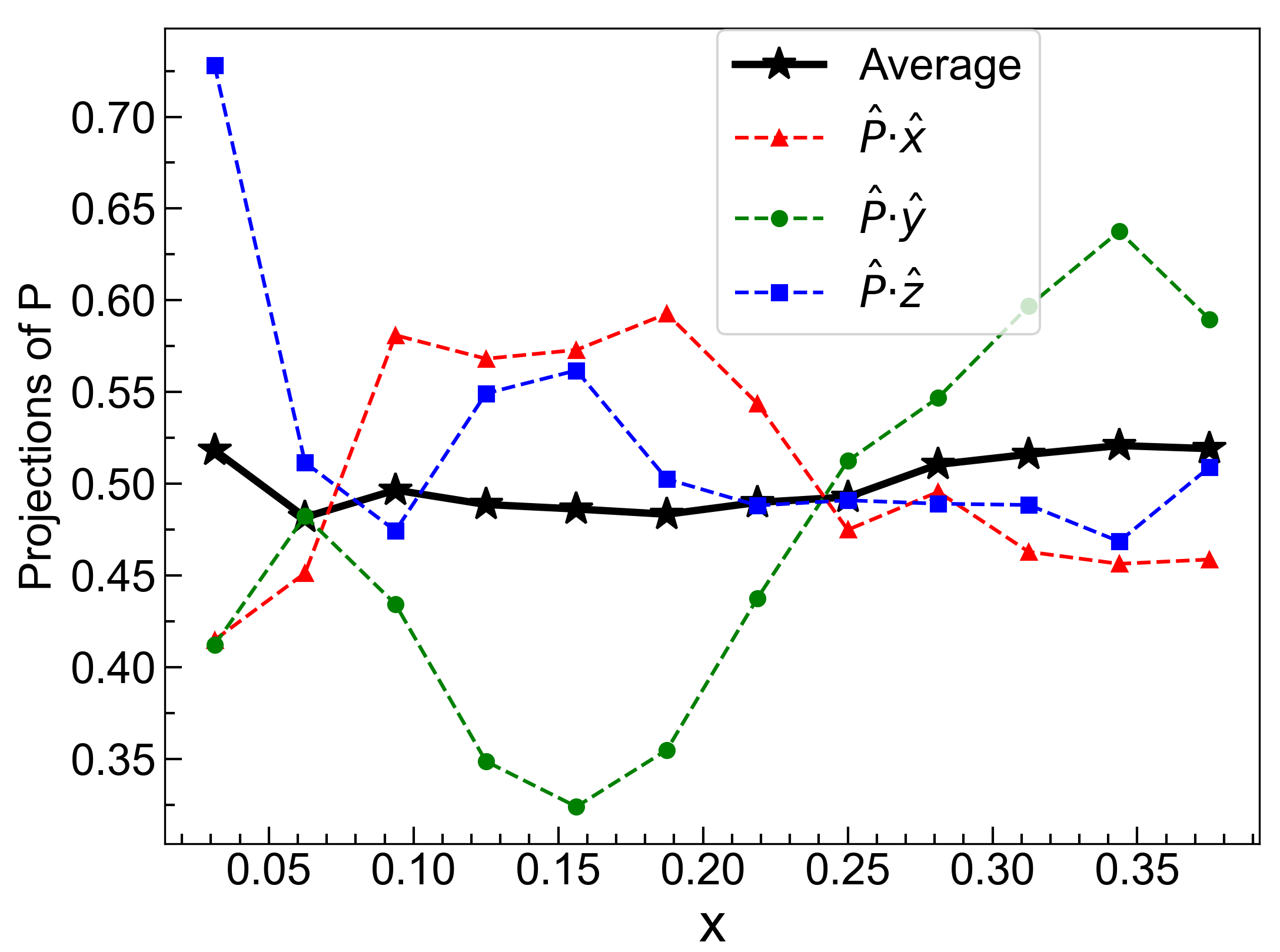}}
\caption{\label{ZT0} Projection of the polarization versor on the Cartesian axes. The average projection corresponds roughly to a [111] direction. }
\end{figure} 

Before proposing an interpretation, we provide info on the direction of {\bf P}. Figure \ref{ZT0} shows the configuration-averaged projections of the polarization versor $\hat{\bf P}$={\bf P}/$P$ on  the Cartesian versors. For this orthorhombic system, this is equivalent to projecting on the crystal axes, numerical factors aside. $\hat{\bf P}$ is parallel to $\hat{\bf y}$ at $x$=0, but of course it fluctuates considerably in the random configurations. Yet, an inspection of the configurations shows  that {\it all} calculated {\bf P} vectors have positive Cartesian components, i.e. lie inside the first Cartesian octant, suggesting a systematic effect related to the underlying structure. Taking a further average of the averaged projections (solid line-and-stars in the Figure), one concludes that $\hat{\bf P}$ is, on average, very roughly parallel  to the (111) crystal direction. 

We suggest that the deviation of ${\bf P}$ from $P$$\hat{y}$ in HfO$_2$, and the rough conservation of $P$ itself, are explained by the underlying structure and local environment around Hf as modified by the substituting V. Hf has seven O neighbors, and V effectively loses one of them (the distance increases from $\sim$2.2 \AA\, to $\sim$2.7 \AA) forming a distorted octahedron with the remaining  six.  Two  V-O bonds dimerize (relative difference  $\sim$15\%) along one of the octahedron's axes (which  point roughly in the direction of the cell diagonals).

Now, in a na\"ive point-charge local-dipoles picture, a longer (oriented) O-to-V bond is a larger dipole that a shorter (oriented) O-to-V bond is, so the dimerization produces a net  dipole pointing along the (oriented) V-to-O short bond. Examining our 9500 V substitutions, we find that all V octahedra are dimerized, and  the configuration-averaged dipole direction points into the first Cartesian octant for all $x$. (Of course, this is not true of all  configurations individually, alhough  only about 1/4 of the bond vector components are negative.) Thus the dimerization distortion of the V octahedron contributes a dipole displacing {\bf P} off the $y$ axis and into the first Cartesian octant. 

Obviously the full polarization change will depend on the rearrangements of all  atoms, and on the changes in dynamical electronic response as measured by the Born effective charges \cite{berry}. We calculate the Born tensors for $n$=2, 5, 12 (lowest energy configurations) to look for trends. Using as proxy the spherical component  of the atom-averaged charge tensor (i.e. one third of its trace), we find an increase of the anomaly of the individual charges vs $x$:  
 $Z_{\rm Hf}$=5.29, 5.37, 5.61;  $Z_{\rm O}$=--2.62, --2.62, --2.69; and  $Z_{\rm V}$=4.51, 4.56, 4.98, for $x$=0.03, 0.15, 0.37 respectively.  Finally, to  measure the average response at different $x$, we use the weighted   cation charge $Z_{\rm c}(x)$=(1--$x$)$Z_{\rm Hf}$+$x Z_{\rm V}(x)$  (the anion charge follows closely the acoustic sum rule $Z_c$+2$Z_a$=0). This is also found to increase with $x$, namely 5.238, 5.241, 5.377 for $x$=0.03, 0.15, 0.37. It thus appears that as $x$ increases the  Born charges of V are sufficiently anomalous to produce an  {\it increasing} overall effective charge anomaly.
 
To summarize our discussion {on polarization}, sustained dynamical charge anomalies and dipolar displacement patterns associated with V make it quite plausible that polarization be conserved in (Hf,V)O$_2$. (This is in line with  \cite{ptov} the  {\it increased}  polarization upon V doping of PbTiO$_3$ and the charge anomalies in V$_2$O$_5$ \cite{v2o5}.)

{We now come to magnetization. Magnetism in this system is entirely due to V being magnetically active as a 4+ ion (one excess electron) and, as discussed in  Sec.\ref{gencons}, to the symmetry of polar HfO$_2$, which  splits the degenerate $d$ states into singly-degenerate states. The latter are essentially all V-like in character, as shown in the following Sec.\ref{secdos}. 
(While this is outside our present scope we note in passing that  symmetry would produce complete degeneracy removal of the $d$ states even, for example, in the non-polar monoclinic ground state of hafnia.)}

{All V configurations are ferromagnetic (or, occasionally, ferrimagnetic as mentioned above). The magnetization per V is $\simeq$1 $\mu_B$, corresponding to  a single unpaired electron. The orbital magnetization contribution is negligible (a factor of 20 smaller than the spin magnetization) as expected from crystal-field angular momentum quenching.}

{To verify that ferromagnetism is the magnetic ground state, we need to investigate "antiferromagnetic"  states. Given disorder and dilution, no conventional antiferromagnetic state (e.g. G-type, A-type, etc.) can be  defined, but we can still build  approximate "AF" states flipping half of the   initial magnetic moments (given the presence of multiple non-equivalent V atoms in random locations, there are many ways to do so). }

{We study several "AF" starting states for the $n$=12, $x$=0.375 case, and find them consistently  disfavored over the FM state. The typical  energy difference is $\delta$E$\simeq$90 meV per cell, which is sizable. This is confirmed by a sample of non-collinear calculations. }

{This difference can be used for a ballpark estimate of the  thermal behavior of the magnetization. Disorder bars the detailed extraction of magnetic couplings, which would be straightforward in ordered systems \cite{danilo}, so we assume arbitrarily a single average coupling and use the standard Ising result for a 2D square lattice (in view of the tendency of V to order in rows and planes, Sec.\ref{structure}) and a 3D cubic lattice with random bimodal disorder \cite{fytas}. The resulting T$_c$'s are 50 K and  64 K respectively.}


\subsection{Density of states and basic optical response}
\label{secdos}

\begin{figure}[ht]
\centerline{\includegraphics[width=1\linewidth]{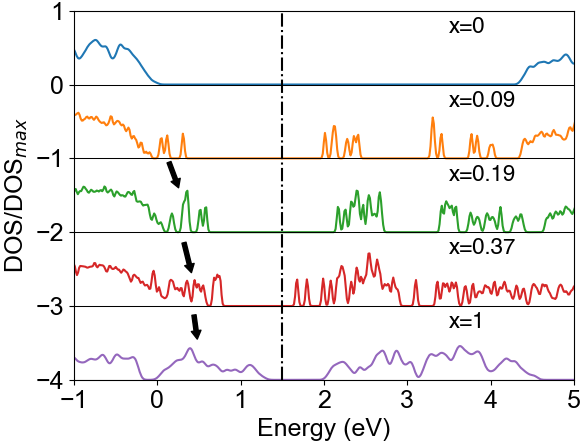}}
\caption{\label{schema-es} {Schematic of the majority-channel density of states vs $x$. Dotted vertical line is the Fermi energy. Arrows sketch the evolution of the V occupied states. Here $x$=1 corresponds to VO$_2$ in its own $C 2/m$ structure.}}
\end{figure}

As discussed in Sec.\ref{gencons}, the origin of the insulating character of the solid solution is a combination of magnetic splitting and simmetry-lowering crystal field splitting. We now attempt a summary of the electronic structure features. 
{Qualitatively, V in HfO$_2$ foreshadows an incipient VO$_2$-like behavior emerging from HfO2. As shown in  Figure \ref{schema-es}, at $x$=0 there is a large, clean gap; at non-zero $x$ a group of V majority states appears just above the valence band, narrow at first but widening as $x$ grows, and separated by a small gap from the V empty states in the upper part of the hafnia gap; in VO$_2$ the lower-lying states are in the same position with respect to the VO$_2$ valence, and the gap is again between mainly V-like states.}

To be more  specific, for $n$ ferromagnetic substituting V, we observe  10$\times$$n$ V $d$ single-electron states in the HfO$_2$ gap,  5$\times$$n$ of which in the majority channel and as many in the  minority. 
Exactly $n$ non-degenerate majority states of V $d$ character accomodate the $n$ V electrons in a  window about 0.5 to 1 eV wide just above the O-derived valence top of HfO$_2$. 
The 4$\times$$n$ empty majority states  are separated by a gap of over 1 eV, decreasing slowly at larger concentrations. In the minority channel, 5$\times$$n$ empty states appear at 2.5-3.5 eV from the occupied states.
All of these features are indeed shared by the same energy region in ferromagnetic VO$_2$.

The magnetic separation, too small in GGA, is restored by GGA+U, which mimicks the selfinteraction cancellation of hybrid functionals \cite{HSE} and selfinteraction-correction \cite{psic}. In fact, even  GGA generally finds degeneracy removal and  insulating character (with tiny gaps).

Inspecting various densities of states (DOS), we  consistently find a gap of roughly 1 to 2 eV (larger at the lowest concentrations and  configuration-dependent). The magnetic splitting is typically over 1 eV larger. Also, note the two lumps of empty V $d$ states separated by about 1.5 eV, possibly a remnant $t_{2g}$--$e_g$ gap. 

\begin{figure}[ht]
\centerline{\includegraphics[width=1\linewidth]{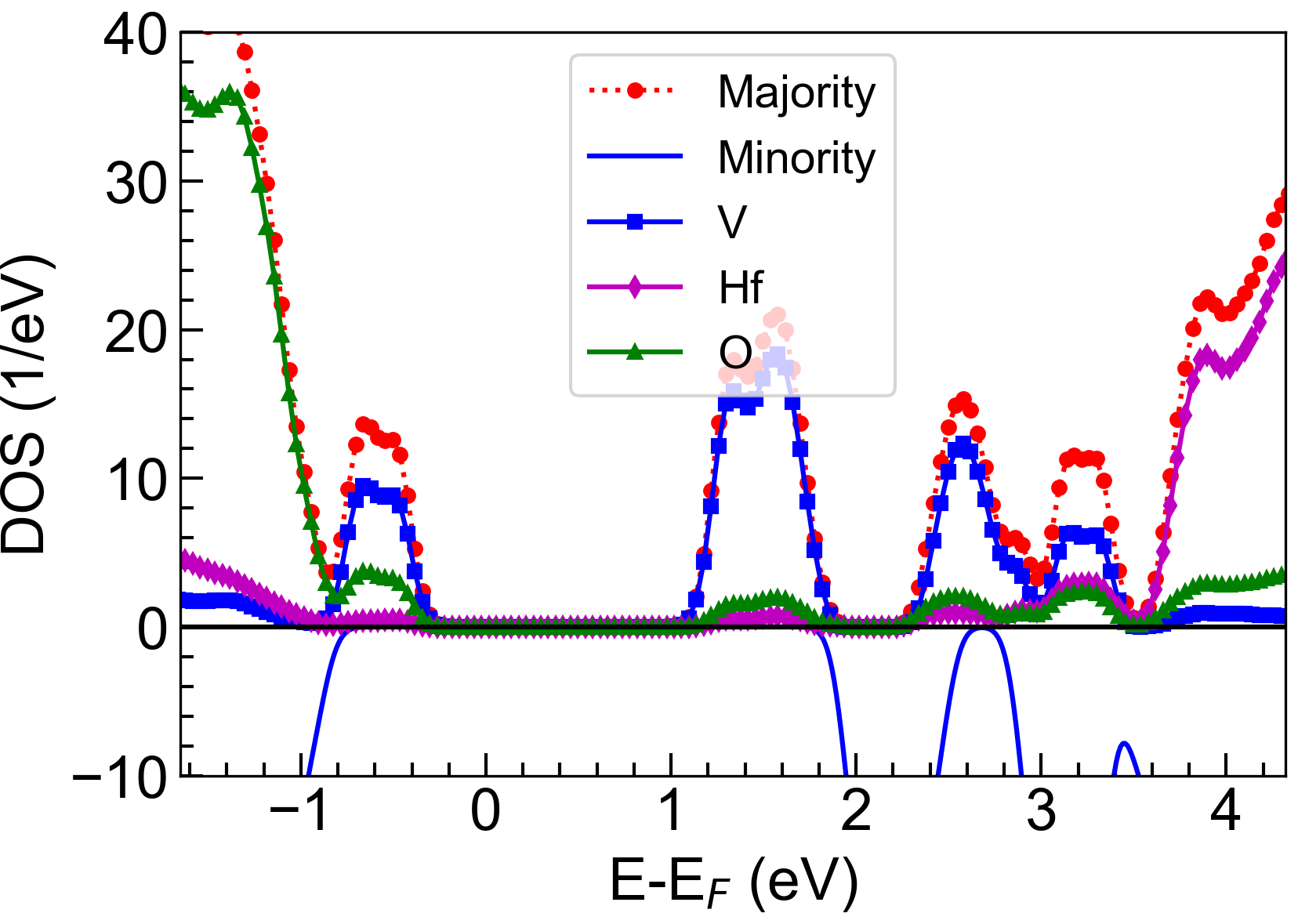}}
\caption{\label{doslarga} Total and atom-projected DOS, $n$=5.}
\end{figure}

For definiteness, we visualize the DOS for the $x$$\simeq$0.15 lowest-energy configuration (in these Figures the Gaussian smearing is 0.1 eV). Figure \ref{doslarga} shows the near-gap region. 
The gap states are of almost pure V and O character, in a roughly 2:1 ratio for the occupied states, larger for the empty ones. 

\begin{figure}[ht]
\centerline{\includegraphics[width=1\linewidth]{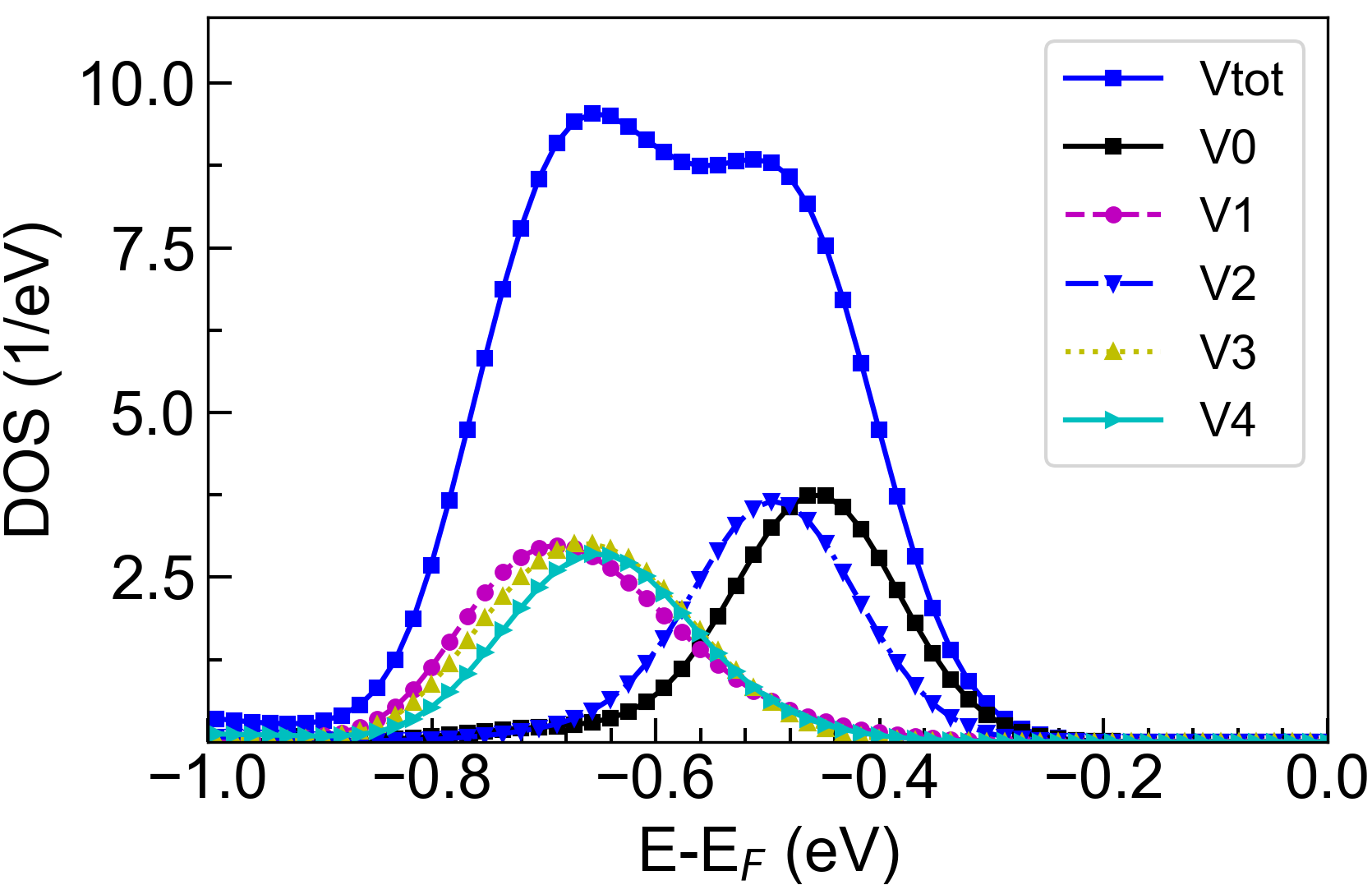}}
\caption{\label{dos8} Projection on V atoms of the occupied majority DOS near E$_F$ for $n$=5.}
\end{figure} 

Figure \ref{dos8} shows  the  DOS of the highest occupied states in the majority channel projected on the V atoms (again for $n$=5, $x$$\simeq$15\%)  summed over the $s$, $p$, and $d$ channels. The  contributions of the five V atoms can be easily made out. These components sum up to produce about two-thirds of the total DOS, the rest coming from oxygen (see Figure \ref{doslarga}).  Analogous results are obtained at other $x$.

\begin{figure}[ht]
\centerline{\includegraphics[width=1\linewidth]{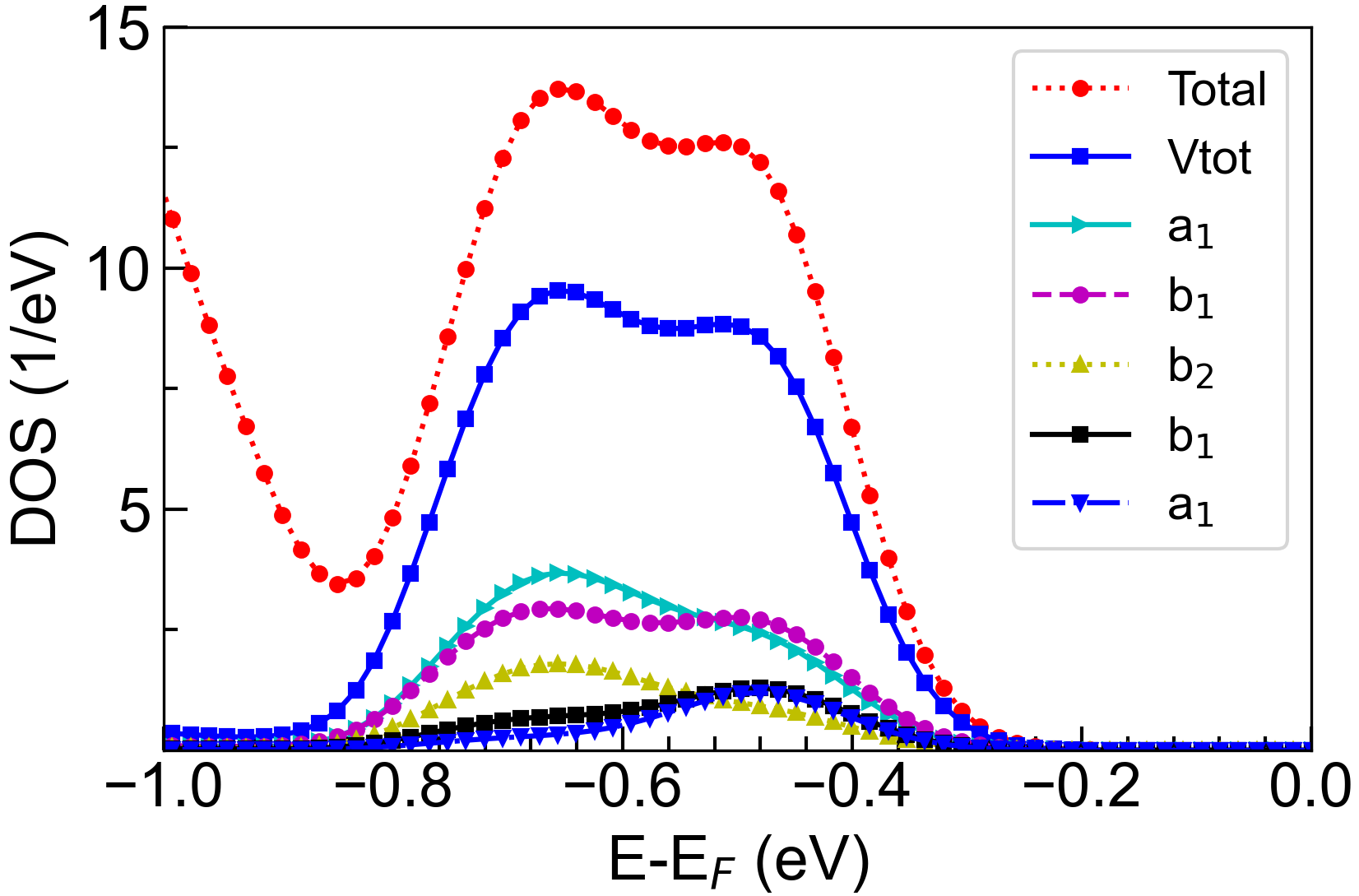}}
\caption{\label{dos2} Orbital-resolved V-projected DOS of the upper occupied  states for $n$=5.}
\end{figure} 

Figure \ref{dos2} shows the orbital character of the occupied V-like gap states, again in the lowest-energy  $n$=5 configuration.  As mentioned, the V weigth is almost totally totally $d$-like, and all the states resulting from the split  $d$ manifold are being admixed. At lower concentration the admixture is similar, though perhaps less marked.

Concerning optical response, dipole-allowed absorption in HfO$_2$ occurs in the UV between occupied O $p$ and empty Hf $d$ bands. As just shown, the occupied states appearing in the HfO$_2$ gap upon V doping have  V $d$ character, which would imply dipole-forbidden transitions, hence no absorption. However, the $mm2$ point symmetry does not conserve parity, so that \cite{dres} certain transitions would be  dipole-allowed depending on light polarization. 
The discussion of these effects  \cite{discpol} would be complicated by the fact that the reference axes vary locally from one V to the other in principle.  Luckily, however, we need not go into it, because symmetry is completely removed in our system due to disorder by random V substitution and attendant local distortions, so that no selection rules apply other than spin conservation. 

\begin{figure}[ht]
\centerline{\includegraphics[width=1\linewidth]{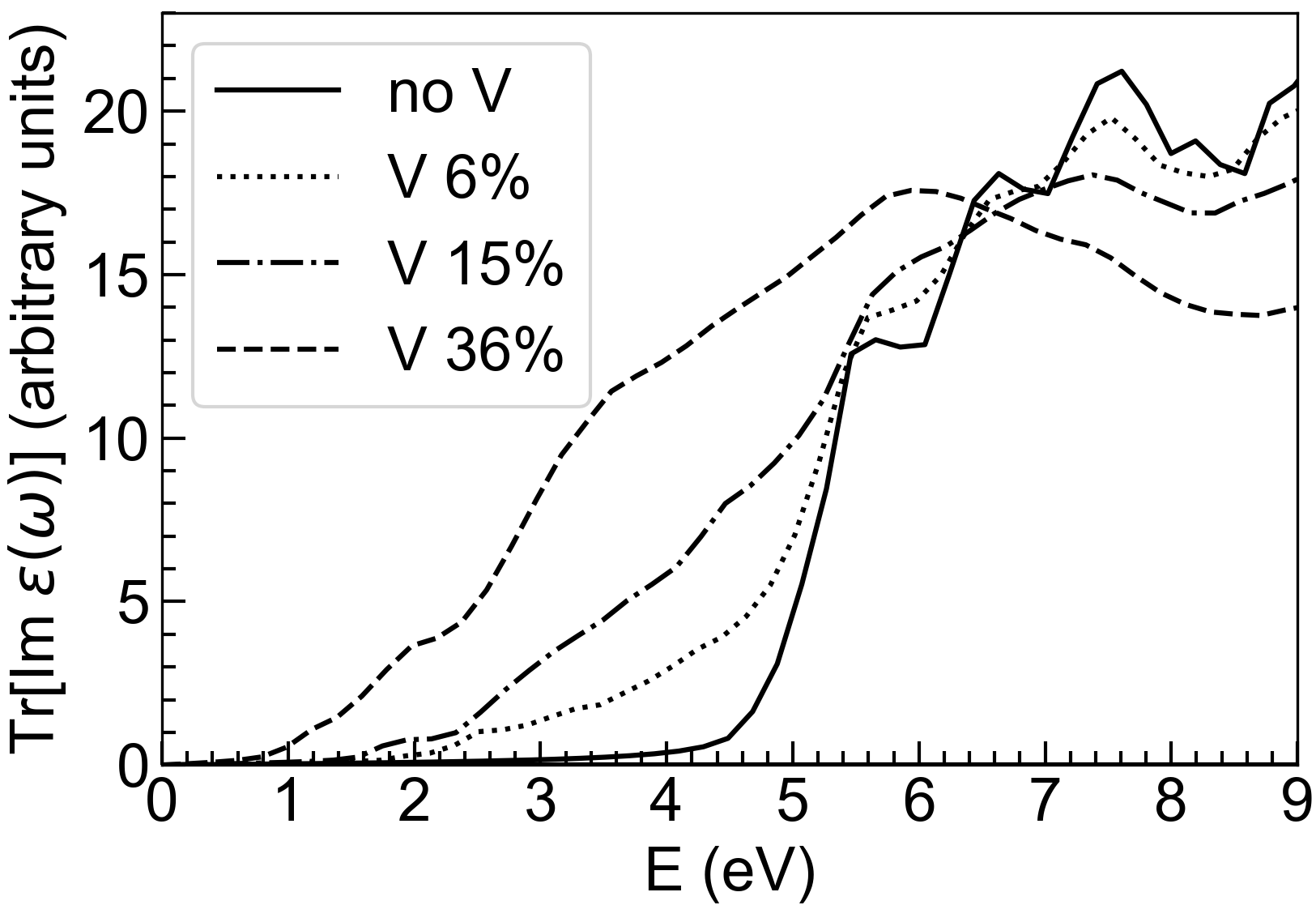}}
\caption{\label{deps} Trace of the imaginary part of the dielecric function tensor at different V concentrations.}
\end{figure}

We can get an idea of the evolution of the absorption spectrum with V concentration from  Figure \ref{deps}, which reports the trace (analogous to a Maxwell-Garnett average) of the imaginary dielectric-function tensor as function of $x$. The absorption onset of hafnia is at the DOS gap of about 4.6 eV as expected. The doped system shows absorption  proportional to V concentration at lower energies: the onsets are at the DOS gaps of 1.9, 1.4, and 0.8 eV for $x$=6, 15, 37\%, and some higher absorptions (clearly visible at 6
\%) appear near 2.5 and 3 eV as expected.

\subsection{Connection to experiment}
\label{expcomp}

There are no experiments on (Hf,V)O$_2$, with  the  important exception of a very recent demonstration of a  metal-ferroelectric-metal stack involving ferroelectric HfO$_2$ doped with V \cite{vhfo2} (we became aware of that work while at an advanced drafting stage of this paper).  Concentrations  in the 3\% to 11\% range were investigated, all of them with analogous ferroelectric properties (and all within our estimated stability range), the optimal  performance being obtained \cite{vhfo2} at 6\% V and  900 K annealing. The temperature of 1000 K we used for  weighting and  entropic contributions is not far from those in \cite{vhfo2}: the atomic-layer deposition growth  is 515 K, and annealing was done between 700 and 1000 K. The immediate substrate layer for the (Hf,V)O$_2$ film is a TiN layer radiofrequency-sputtered on silica, presumably disordered. Our free-standing simulated cell may then be a not too inappropriate model for the hafnia overlayer.

\begin{figure}[ht]
\centerline{\includegraphics[width=1\linewidth]{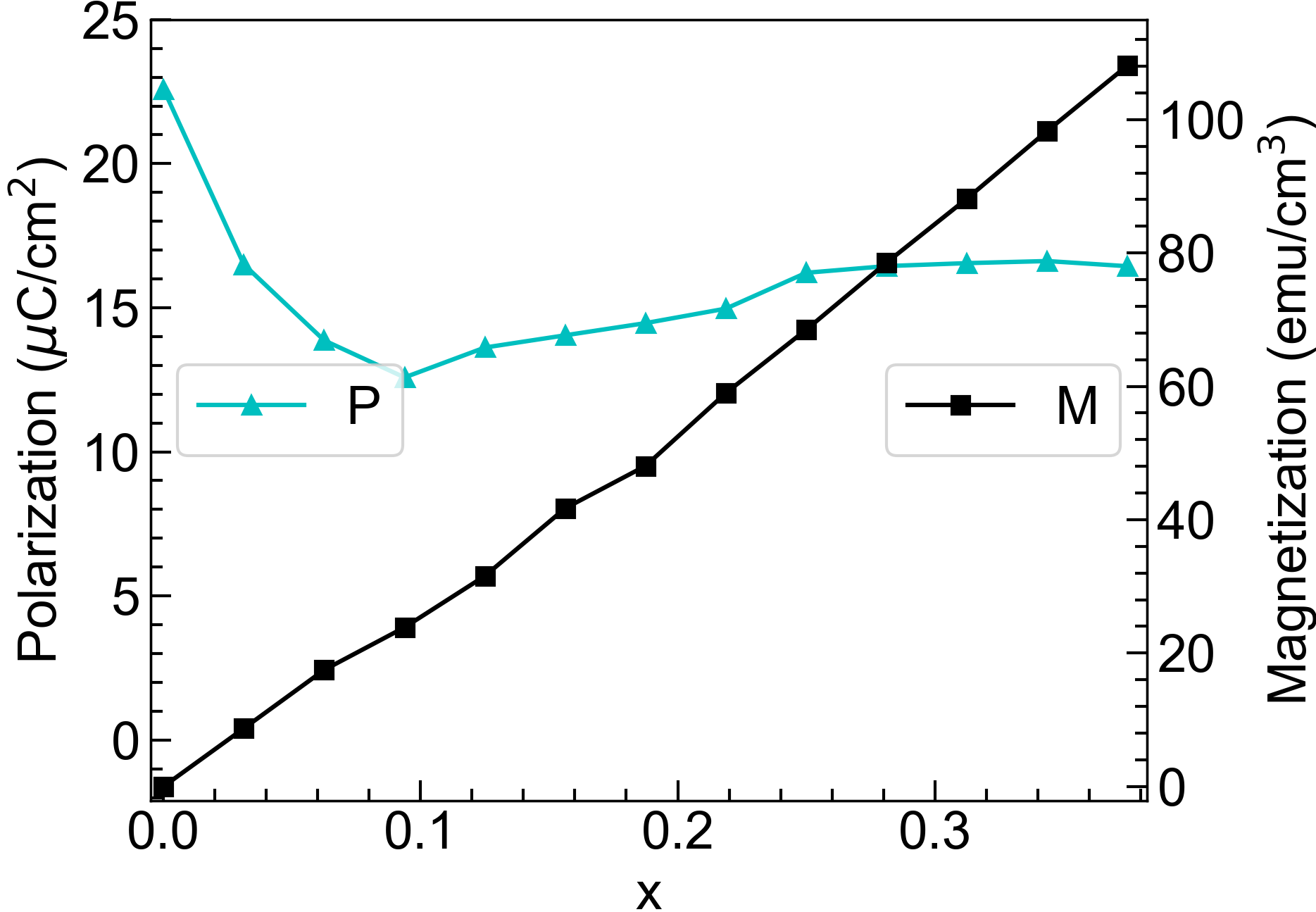}}
\caption{\label{MPunits} $P$ (left $y$ axis) and total $M$ (right $y$ axis) in human units vs. concentration.}
\end{figure} 

The typical remnant polarization during testing was between 15 and 25 $\mu$C/cm$^2$, as was the saturation polarization. This is in quite decent agreement with the typical polarization predicted here, as can be seen in Figure \ref{MPunits}, which reports polarization and magnetization in standard units.
The predicted magnetization vs $x$ (also in Figure \ref{MPunits}) is  $M$($x$)$\simeq$290 $x$ emu/cm$^3$, so 
at the  predicted stability limit of 16\%, $M$$\simeq$46 emu/cm$^3$, and at the 6\% V adopted in \cite{vhfo2} $M$ is a still respectable 17 emu/cm$^3$. 

This prediction applies to a state in which all V is acting as a 4+ ion. The XPS performed on samples at 6\% and 11\% V  suggest that about two thirds (estimated from Fig.S1 of \cite{vhfo2}) of the V atoms have 3+ valence, and one third have  4+. Valence 3+  corresponds to V$_2$O$_3$ local composition, and involves the formation of O vacancies in the HfO$_2$ matrix to mantain neutrality and insulating character. (This is indeed the only possible route  in a dioxide when the substituting cation is purely 3+,  as found long ago for Y-stabilized zirconia \cite{yzro} and  hafnia-alumina mixtures \cite{hfal2o3}.) A description of the  mixed valence system is outside our present scope, but our results prove that  V can act as a 4+ ion and still produce an insulating and ferroelectric state in HfO$_2$ without compensating defect such as vacancies, which explains the otherwise puzzling   one-third population of 4+ V ions.

Notably, the V 4+ ion is magnetic in the dioxide whereas the V 3+,  in the local sesquioxide stoichiometry facilitated  by O vacancies, \textcolor{red}{may or may not be. If it is not,} assuming the ratio of 3+ to 4+ V is indeed 2:1 as inferred from XPS, the samples of Ref.\cite{vhfo2} with 3\% to 11\% V should already exhibit a magnetization of 3 to 11 emu/cm$^3$. Further,  an increase of the oxygen partial pressure (O-rich conditions) during growth will decrease the concentration of O vacancies, hence increase  the fraction of magnetic 4+ V ions and the magnetization. {Importantly, some oxygen deficiency appears to stabilize the hafnia polar phase; V helps this stabilization with its multivalence behavior}.  It barely needs  specifying that no action is, or needs to be, taken to enforce a specific charge state of the V ion, which adapts automatically to the chemical environment. \\

\subsection{Structure}
\label{structure}

\begin{figure*}[ht]
\centerline{\includegraphics[width=0.4\linewidth]{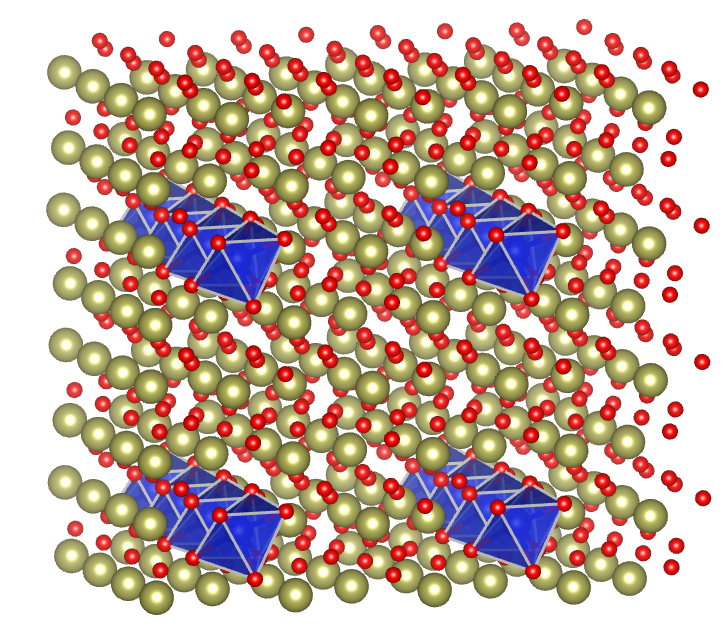}
\includegraphics[width=0.4\linewidth]{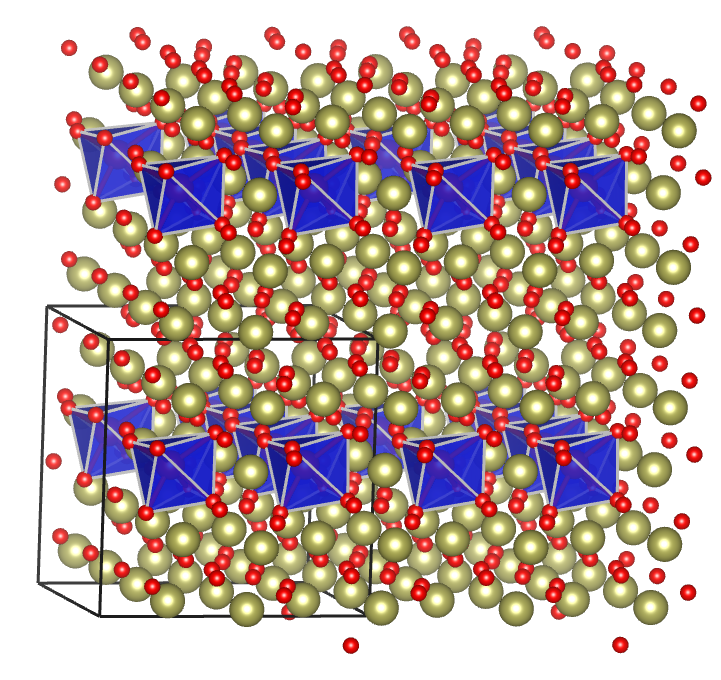}}
\centerline{\includegraphics[width=0.4\linewidth]{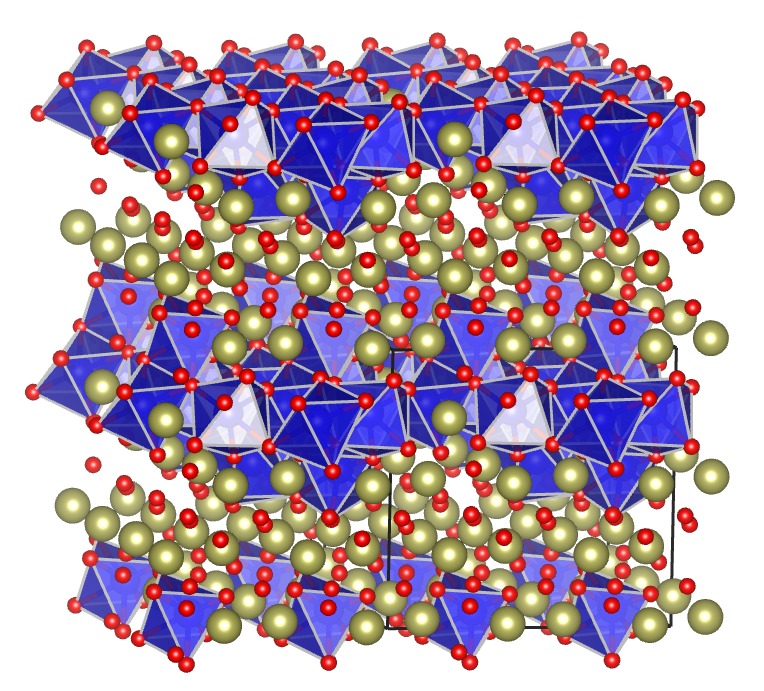}\includegraphics[width=0.4\linewidth]{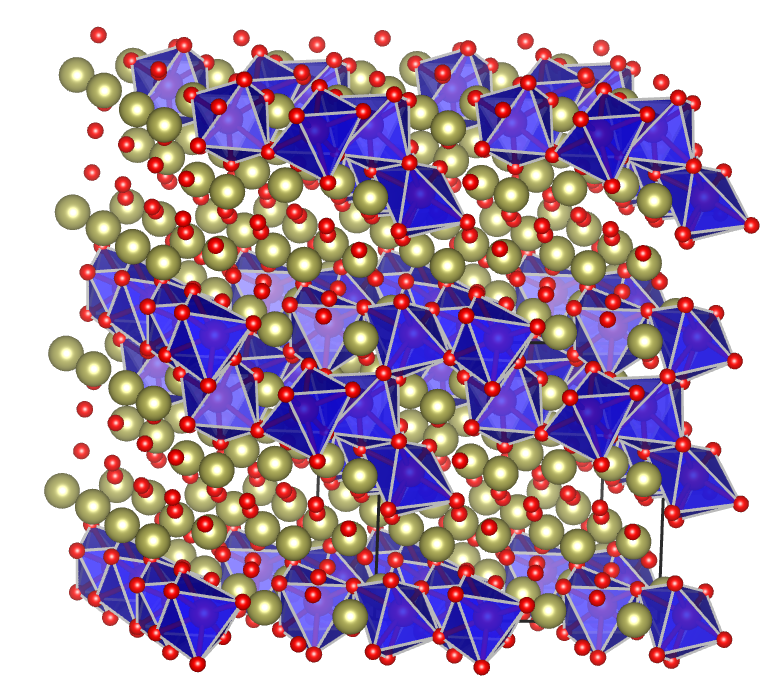}}
\caption{\label{strutturem} Minimum-energy structures for $n$=2, 3, 9, 11 (clockwise from top left), viewed from a direction in the first octant. The vertical direction is $y$.}
\end{figure*}

\begin{figure*}[ht]
\centerline{\includegraphics[width=0.4\linewidth]{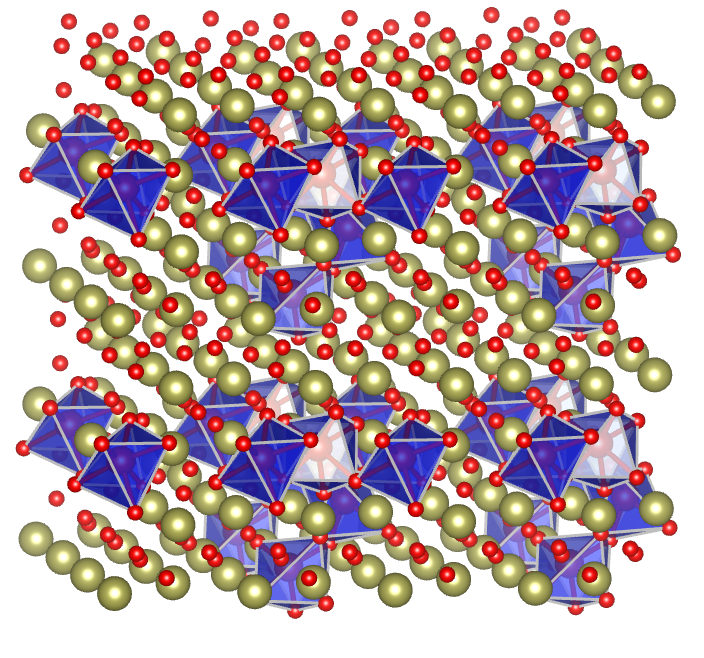}\includegraphics[width=0.44\linewidth]{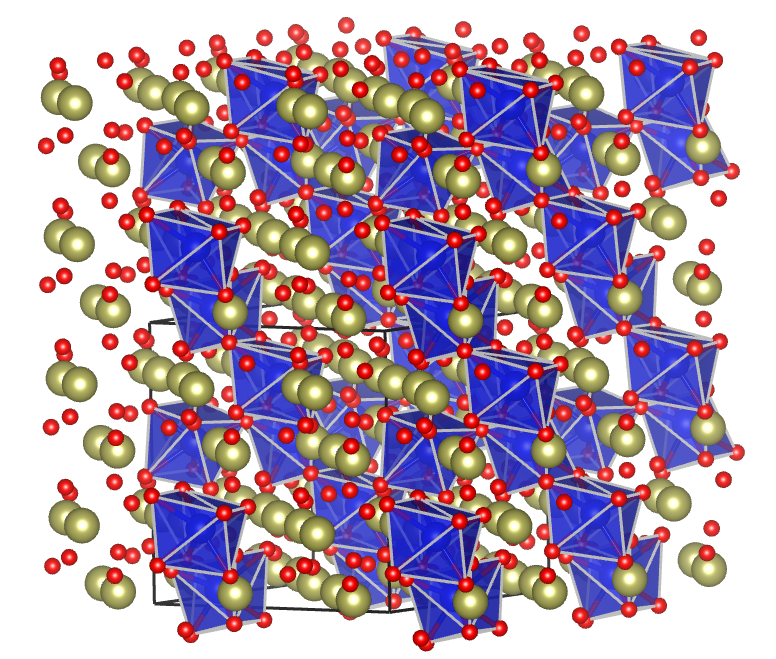}}
\centerline{\includegraphics[width=0.4\linewidth]{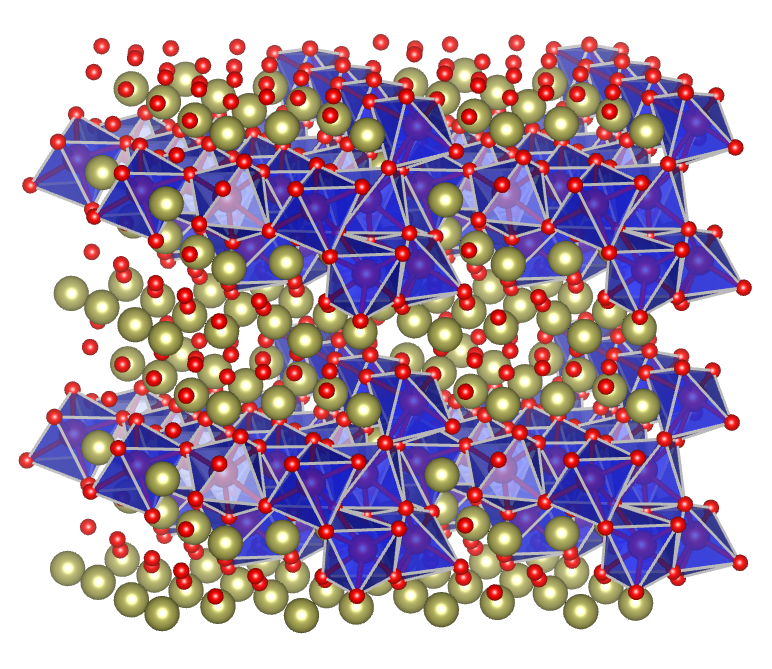}\includegraphics[width=0.415\linewidth]{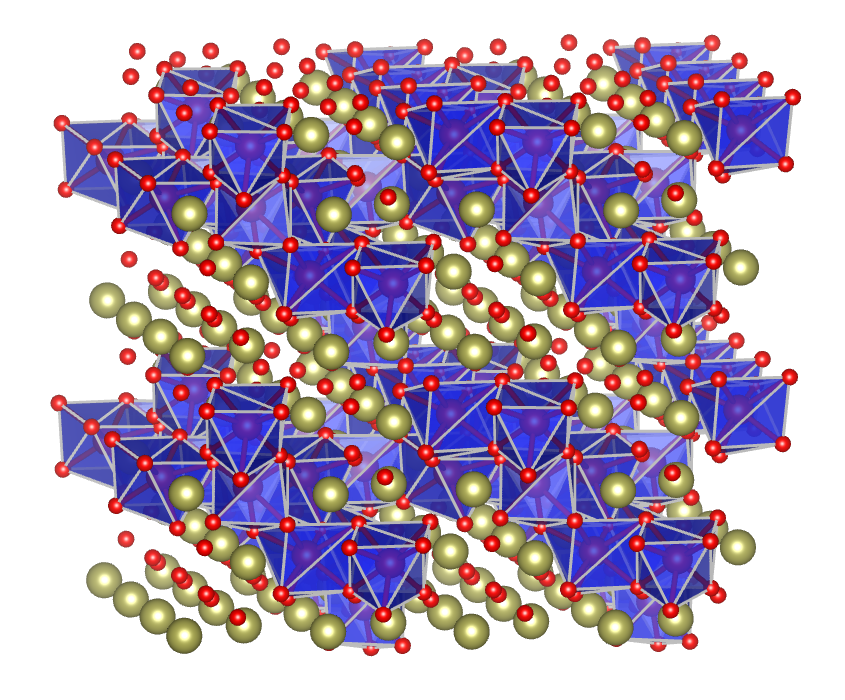}}
\caption{\label{struttureM} Minimum-energy (left) vs maximum-energy (right) structures for $n$=5 (top) and $n$=12 (bottom). Same view as the previous Figure. The vertical direction is $y$.}
\end{figure*}

The large number of configurations makes an analysis of  possible structural patterns  quite difficult. We elect to inspect  some of the low-energy structures for a few concentrations. The result is hard to convey visually, but 
analyzing the cells in 3D it is quite evident that V has a tendency, as concentration increases, to  order first in rows (a trend noted earlier in \cite{vlto}), then in multiple rows leading to incomplete or almost complete planes, then to multiple incomplete and "decorated" planes, which tend to  stack along the {\bf b} and (to a lesser extent) the {\bf a} crystal axes. We attempt a summary in Figures \ref{strutturem} and \ref{struttureM}, showing 2$\times$2$\times$2 replicas of the simulation cell from a point of view in the first Cartesian octant near the $xy$ plane. The vertical axis is the $y$ direction. Only V-centered polyhedra  are shown.

Figure \ref{strutturem} displays  the lowest energy structures for $n$=2, 3, 9 and 11. Clearly the structures mutate from rows ($n$=2) to intersecting rows forming largely incomplete planes ($n$=3), towards incomplete, decorated planes ($n$=9) to practically complete planes with decorations above and below ($n$=11, as well as $n$=12 in Figure \ref{struttureM}). For the latter two, for example, there are seven V's in one plane with 8 available cation sites per cell, which is a rather good guess at a "layered segregated" structure considering the totally random process leading to it and the fact that our random-sample size is $\sim$100 out of a binomial configurational spaces of size ranging from many thousands to  hundred millions for our concentrations.  

Finally, in Figure \ref{struttureM} we  compare the highest-energy structures  with the lowest-energy ones in the cases $n$=5 and $n$=12. The differences again are that V clustering is less compact in the high-energy cases, while  almost complete planes tend to emerge in the low-energy one.

\section{Summary and acknowledgments}

Based on extensive ab initio calculations we have shown that  ferroelectric HfO$_2$ mixed with vanadium up to 30\% conserves a major fraction ($\sim$70\%) of its polarization while acquiring a respectable ferromagnetic magnetization  proportionally to V content ($\sim$2.9 emu/cm$^3$ per V \%). The mixing free energy indicates phase stability against the binary compounds (in the respective structures) up to around 16\% V. Good agreement is found with a very recent experiment. (Hf,V)O$_2$ then emerges as a robust ferromagnetic multiferroic with substantial magnetization and polarization, and thus promising for applications. The data supporting the findings above are openly available \cite{dati}.

We thank Andrea Urru, Niccol\`o Martinolli, and Alessio Filippetti for a critical reading; CINECA Bologna for the ISCRA Supercomputing Grant {\tt MAFNIA2} on EuroHPC Leonardo; {RES-BSC Barcelona Supercomputing Center for supercomputing time on MareNostrum5 GPP}. VF acknowledges Universit\`a di Cagliari for a sabbatical leave and TU Dresden for a Senior DRESDEN Fellowship. VF and PA dedicate this paper to the memory of dr. Silvia Piacentini.

\begin{thebibliography}{References}
\bibitem{convex}
For $x$$>$0, a function $x^a$ with 0$<$$a$$<$1 is upward convex and monotonically increasing, whereas if 1$<$$a$ it is downward convex.
020).
\end{thebibliography}
\end{document}